\documentclass[sn-mathphys,Numbered]{sn-jnl}% Math and Physical Sciences Reference Style
\usepackage{subfigure}
\usepackage{hyperref}

\usepackage{graphicx}%
\usepackage{multirow}%
\usepackage{amsmath,amssymb,amsfonts}%
\usepackage{amsthm}%
\usepackage{mathrsfs}%
\usepackage[title]{appendix}%
\usepackage{xcolor}%
\usepackage{textcomp}%
\usepackage{manyfoot}%
\usepackage{booktabs}%
\usepackage{algorithm}%
\usepackage{algorithmicx}%
\usepackage{algpseudocode}%
\usepackage{listings}%
\theoremstyle{thmstyleone}%
\theoremstyle{thmstyletwo}%
\theoremstyle{thmstylethree}%

\begin{document}

\title[Article Title]{Muon/Pion Identification at BESIII based on Variational Quantum Classifier}


\author[1]{\fnm{Zhipeng} \sur{Yao}}
\author[1]{\fnm{Xingtao} \sur{Huang}}
\author*[1]{\fnm{Teng} \sur{Li}}\email{tengli@sdu.edu.cn}
\author[2]{\fnm{Weidong} \sur{Li}}
\author[2]{\fnm{Tao} \sur{Lin}}
\author[2]{\fnm{Jiaheng} \sur{Zou}}

\affil[1]{Shandong University, Qingdao, Shandong, 266237, People’s Republic of China}

\affil[2]{Institute of High Energy Physics, Beijing, 100049, People’s Republic of China}


\abstract{In collider physics experiments, particle identification (PID), i. e. the identification of the charged particle species in the detector is usually one of the most crucial tools in data analysis. In the past decade, machine learning techniques have gradually become one of the mainstream methods in PID, usually providing superior discrimination power compared to classical algorithms. In recent years, quantum machine learning (QML) has bridged the traditional machine learning and the quantum computing techniques, providing further improvement potential for traditional machine learning models. In this work, targeting at the $\mu^{\pm}  /\pi^{\pm}$ discrimination problem at the BESIII experiment, we developed a variational quantum classifier (VQC) with nine qubits. Using the IBM quantum simulator, we studied various encoding circuits and variational ansatzes to explore their performance. Classical optimizers are able to minimize the loss function in quantum-classical hybrid models effectively. A comparison of VQC with the traditional multiple layer perception neural network reveals they perform similarly on the same datasets. This illustrates the feasibility to apply quantum machine learning to data analysis in collider physics experiments in the future.}

\keywords{Particle identification, Quantum Machine Learning, Variational Quantum Classifier, BESIII Experiment}



\maketitle

\section{Introduction}\label{sec1}

The Beijing Spectrometer III (BESIII) is a collider physics experiment running on the Beijing Electron–Positron Collider II (BEPC II) \cite{bib1}. BESIII is designed to study the physics of tau, charm, charmonium, and light hadron decays, as well as to search for new physics beyond the Standard Model. The particle identification (PID), i. e. the identification of charged particle species, is one of the most fundamental tasks in various BESIII physics studies, such as the precise $f_{D}/f_{D_{s}}$ measurements as well as precise measurements of CKM elements $V_{cs}$ and $V_{cd}$. 

In the past two decades, machine learning techniques have armed high energy physics (HEP) experiments with a powerful toolbox \cite{radovic2018machine,feickert2021living}. Frequently used models include the boosted decision tree (BDT) \cite{bib2}, the artificial neural network \cite{bib3}, and support vector machine (SVM), etc. In the domain of PID, machine learning models are proved to be often superior to traditional algorithms, owing to the capability of extracting useful features among a large number of input features. However, with the development of future HEP experiments, current PID models are faced with higher challenges. The next-generation collider physics experiments, such as the super tau charm facility (STCF), require better PID performance over a much wider momentum range, requiring much more advanced PID technologies.

In recent years, the emergence of quantum computing has offered new possibilities for HEP experiments. In particular, the development of quantum machine learning (QML) techniques \cite{bib4,bib5} has bridged the quantum computing and the machine learning domains, and has provided insights for solving problems in HEP \cite{guan2021quantum,wu2022challenges}. At present, commonly used QML models for classification problems include the quantum support vector machine (QSVM) \cite{bib6} and the variational quantum classifier (VQC) \cite{bib7} etc. A few previous studies \cite{bib8,bib9} have shown that these QML models have the potential to outperform traditional machine learning models in discrimination performance when dealing with complex physical data. Therefore, at this stage, it is useful and interesting to study the feasibility of applying QML models to various data processing tasks at HEP experiments. 

In this paper, we introduce a VQC model developed to solve the $\mu^{\pm}  /\pi^{\pm}$ identification problem at BESIII. By testing the model on noisy intermediate-scale quantum (NISQ) simulators and comparing the performance with traditional machine learning algorithms, we explore the feasibility of applying QML to PID problems. The basic working principles of VQC are introduced in Sect. \ref{sec2}. Then in Sect. \ref{sec3}, the PID system of the BESIII detector and the dataset used for training and testing are briefly summarised. Finally, we show the performance of the VQC model as well as the comparison with the traditional machine learning methods in Sect. \ref{sec4}.

\section{Variational Quantum Classifier}\label{sec2}

The VQC is a classical-quantum hybrid model developed to solve classification problems based on the variational quantum algorithm. The working principle of VQC is similar to that of classical machine learning models for classification problems, which can be summarised into the following four parts.

\begin{enumerate}
\renewcommand{\labelenumi}{(\theenumi)}
    \item Prepare the input data vectors $\vec{x}$.
    \item Construct a suitable model $f(\vec{x} ,\boldsymbol{\theta})$ to fit the data $\vec{x}$, where $\boldsymbol{\theta}$ denotes the parameters to be optimized during the training process.
    \item Construct a loss function $L(\vec{x} ,\boldsymbol{\theta})$ to quantify the difference between the true and predicted labels, indicating the degree of fitting between the model and the data.
    \item Use some optimizers to find the minimum value of $L(\vec{x} ,\boldsymbol{\theta})$ to determine the optimal parameters $\boldsymbol{\theta}$ that minimize $L(\vec{x} ,\boldsymbol{\theta})$.
\end{enumerate}

\begin{figure}[h]
\centering
\includegraphics[scale=0.46]{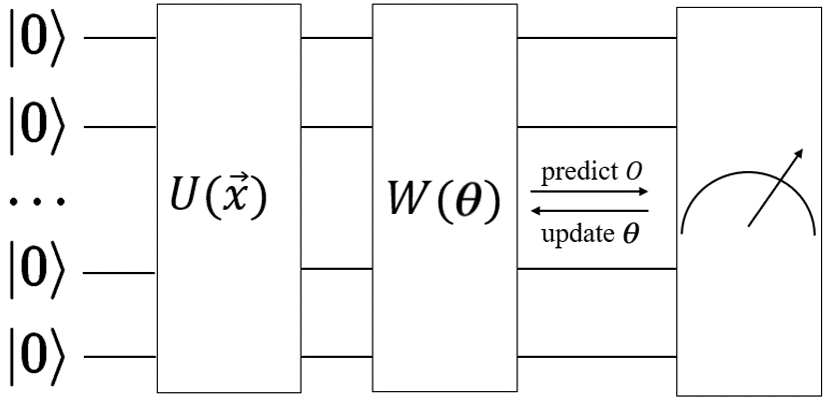}
\caption{\centering Variational Quantum Classifier}
\label{Fig1}
\end{figure}

In general, VQC (as shown in Fig. \ref{Fig1}) follows the same process as described above with some changes to fit in the quantum environment. Firstly, the training data is encoded into quantum states by a feature map $U_{(\vec{x})}$ and then processed in high-dimensional Hilbert spaces. Secondly, the model $f(\vec{x} ,\boldsymbol{\theta})$ is built on the variational quantum circuit $W(\boldsymbol{\theta})$ composed of quantum logic gates. Lastly, the gradient is computed based on the measurement outcomes of the quantum circuit, enabling classical optimizers to adjust the parameters and minimize the loss function, thereby enhancing the performance of the model.

\subsection{Data Encoding Circuit}

The data encoding, as the crucial start of the VQC, provides a way to re-express the input data as quantum states. This step is performed by the encoding circuit that projects the input data vector $\vec{x}$ onto a quantum state $\left\lvert\psi_{in}(\vec{x})\right\rangle$ described in a high-dimensional Hilbert space. The data encoding we applied in this study is proposed in \cite{bib10} formally defined as: 
\begin{equation}
\mathcal{U}_{(\vec{x})}=U_{(\vec{x})} H^{\otimes n} \cdots U_{(\vec{x})} H^{\otimes n}\label{eq1}
\end{equation}
In Eq. \eqref{eq1}, $H$ is the conventional Hadamard gate and $U_{(\vec{x})}$ is a unitary gate. $H$ and $U_{(\vec{x})}$ can repeat multiple times to enhance the expressibility. The circuit consisting of these gates acts on the $n$ ground states $\left \lvert 0  \right \rangle ^{n}$. Because the matrix representation of the $H$ gate:
\begin{equation}
    H=\frac{1}{\sqrt{2}}\left(\begin{array}{cc}
1 & 1 \\
1 & -1
\end{array}\right)
\end{equation}
is a $\pi$ rotation about the axis that bisects the angle between the $X$-axis and the $Z$-axis. It has the effect of changing the computation basis from $\left \lvert 0  \right \rangle$, $\left \lvert 1  \right \rangle$ to a superposition state $\left \lvert + \right \rangle $, $\left \lvert - \right \rangle $ and vice-versa. In our study, we use the Pauli Expansion circuit $U_{(\vec{x})}$:
\begin{equation}
    U_{(\vec{x})}=\exp \left(i \sum_{S \subseteq[n]} \phi_{S}(\vec{x}) \prod_{i \in S} P_{i}\right)
\end{equation}
The $P_i \in \left \{ I,X,Y,Z \right \}$ represent the Pauli matrices while $S$ denotes the connectivities of qubits.
And  $\phi_{S}$ denotes the data mapping function: 
\begin{equation}
    \phi_{S}(\vec{x})=\left\{\begin{array}{cl}
x_{i} & \text { if } S=\left \{ i \right \} \\
\prod_{j \in S}\left(\pi-x_{j}\right) & \text { otherwise }
\end{array}\right.
\end{equation}
In this study, for simplicity, only one-qubit and two-qubit logical operations are used to construct feature maps. The feature $x_i$ is scaled to the [0, 2] range as an angle before encoding.  Based on the basic structures, several different feature maps are simulated to search for the optimal configuration. The performance of these feature maps is introduced later. 
\subsection{Variational Circuit}
The variational circuit $W(\boldsymbol{\theta})$ acts on the quantum state $\left\lvert\psi_{i n}(\vec{x})\right\rangle$ prepared earlier and generates output states $\left\lvert\psi_{\text {out }}(\vec{x})\right\rangle$. The variational circuit contains free parameters to be trained by a classical optimization algorithm.
\begin{equation}
    \begin{split}
    \left\lvert\psi_{\text {out }}(\vec{x})\right\rangle &= W(\boldsymbol{\theta})\left\lvert\psi_{\text {in }}(\vec{x})\right\rangle  \\
        &=U_{\text {rot }}(\theta_i) U_{\text {ent }} \ldots U_{\text {rot }}(\theta_j) U_{\text {ent }}\left\lvert\psi_{i n}(\vec{x})\right\rangle
    \end{split}\label{eq6}
\end{equation}

Eq. \eqref{eq6} formally defines the variational circuit, whose free parameters define $U_{\mathrm{rot}}(\boldsymbol{\theta})$, consisting of a series of rotation on qubits about the $X, Y, Z$ axis. For example, single-qubit rotations about the $X, Y, Z$ axis used in our study are defined as:
\begin{equation}
    R X(\theta)=\exp \left(-i \frac{\theta}{2} X\right)=\left(\begin{array}{cc}
\cos \frac{\theta}{2} & -i \sin \frac{\theta}{2} \\
-i \sin \frac{\theta}{2} & \cos \frac{\theta}{2}
\end{array}\right)
\end{equation}
\begin{equation}
    R Y(\theta)=\exp \left(-i \frac{\theta}{2} Y\right)=\left(\begin{array}{cc}
\cos \frac{\theta}{2} & -\sin \frac{\theta}{2} \\
\sin \frac{\theta}{2} & \cos \frac{\theta}{2}
\end{array}\right)\label{eq8}
\end{equation}
\begin{equation}
    R Z(\theta)=\exp \left(-i \frac{\theta}{2} Z\right)=\left(\begin{array}{cc}
e^{-i \frac{\theta}{2}} & 0 \\
0 & e^{i \frac{\theta}{2}}
\end{array}\right)
\end{equation}
$U_{ent} $ in Eq. \eqref{eq6} refers to the entanglement between different qubits. Similarly, we can apply $U_{\mathrm{rot}}(\boldsymbol{\theta})$ and $U_{ent} $ multiple times to adjust the number of repetitions of the circuit, thus using more parameters to enhance the complexity of the circuit to make the circuit better fit the data. 

\subsection{Measurement}
The output (or classification score) of VQC depends on the result of the measurement of qubits that define an observable $\widehat{O}$. VQC interprets the measured bitstring as the output of a classifier by applying a parity mapping.

According to the basic principle of quantum mechanics, the expectation value for measuring the observable $\widehat{O}$ is deterministic. The probability distribution of the state of the measured qubits can be estimated by calculating the weighted average after repeating the experiment a reasonable number of times. The number of repetitions defines shots for the VQC model and is set to 1024 in this study. The general form of the expectation value, as the output of the VQC is given by
\begin{equation}
    f(\boldsymbol{\theta})=\left \langle \psi  \lvert W^{\dagger }(\boldsymbol{\theta}) \widehat{O}W(\boldsymbol{\theta}) \lvert\psi  \right \rangle
\end{equation}
The main objective of the VQC is to determine an appropriate $\boldsymbol{\theta}$ that effectively captures the underlying patterns in the data, ensuring that the predictions align with the true labels. This is accomplished by evaluating the fit between the model and the data using a loss function $L(\vec{x},\boldsymbol{\theta})$, and subsequently minimizing this loss function.
The cross-entropy \cite{bib11}, which is commonly used for classification tasks, is employed as the loss function for VQC.

\subsection{Optimization}\label{sec2.4}
The most critical part of the VQC is the optimization of the variational circuit parameterized by $\boldsymbol{\theta}$ that must minimize the loss function. At present, the general optimization process in the field of machine learning is carried out by the gradient descent algorithm. 
\begin{equation}
    \boldsymbol{\theta} _{n+1} = \boldsymbol{\theta} _{n} - \eta  f{}'(\boldsymbol{\theta} _{n} ) \label{eq13}
\end{equation}
Eq. \eqref{eq13} indicates how the parameters are optimized in a recursive procedure. The $\eta$ denotes the learning rate, while the $ f{}'(\boldsymbol{\theta} _{n} )$ denotes the gradient computed through back-propagation. Back-propagation is a technique in the reverse mode of the automatic differentiation algorithm, which propagates the error between the output and the true label from the output layer to the input layer. It then calculates the contribution of each parameter to the error using the chain rule, allowing for parameter updates to minimize the loss function.

As one step forward, Newton's method makes use of the second-order derivative when updating the parameters, as shown in Eq. \eqref{eq14}.
\begin{equation}
    \boldsymbol{\theta} _{n+1} = \boldsymbol{\theta} _{n} - \frac{ f{}'(\boldsymbol{\theta} _{n} )}{f{}''(\boldsymbol{\theta} _{n}) } \label{eq14}
\end{equation}
Newton's method is based on a linear transformation of the Hessian matrix $H$ (second-order derivative) over the gradient for the search direction so it usually converges faster. But Newton's method is usually time-consuming on large datasets given it needs to invert the Hessian matrix. To mitigate this, an n-order matrix $G_{n} $ can be used to approximate $H^{-1}$, as the basic idea of the quasi-Newton method \cite{davidon1991variable}. 

At present, both gradient descent and the quasi-Newton method play vital roles in the field of quantum machine learning, yet it is challenging to compute gradients on quantum devices due to the inability to save intermediate results. Previously, there was a tendency for variational circuits such as variational quantum eigensolver (VQE) to use gradient-free methods, mathematically known as numerical differentiation
\begin{equation}
    \frac{\mathrm{d} f(\boldsymbol{\theta} )}{\mathrm{d} \boldsymbol{\theta} } \approx\frac{f(\boldsymbol{\theta} +h)-f(\boldsymbol{\theta} -h)}{2 h}+O\left(h^{2}\right)\label{eq15}
\end{equation}
as an optimization strategy. In Eq. \eqref{eq15} $h$ is a sufficiently small value and $\frac{\mathrm{d} f(\boldsymbol{\theta} )}{\mathrm{d} \boldsymbol{\theta} }$ has an truncation error $O\left(h^{2}\right)$. As the parameter space becomes larger, the excessive number of epochs leads to a heavy accumulation of errors.

Given the advantages of automatic differentiation for machine learning \cite{bib12}, there are now software frameworks such as $PennyLane$ that have extended automatic differentiation to variational quantum circuits \cite{bib13}. Premised on the expectation of observable measurements as output, it designs quantum nodes, together with classical nodes that can contain numerical calculations, to form a computational graph, and then processes the gradient via automatic differentiation.

Currently, the gradient is more popularly computed based on the $parameter$ $shift$ $rule$ proposed by Schuld et al \cite{bib14}. It is an analytic gradient obtained by a rigorous derivation of the formula, so it possesses higher accuracy compared with numerical calculation and is generally accepted in practice. If the Hermitian generator of a parameterized quantum gate in the variational circuit has two eigenvalues, then the $parameter$ $shift$ $rule$ can find the analytical solution of the gradient by two quantum circuits with shifted macroscopic parameters. Take $RY(\theta )$ in Eq. \eqref{eq8} as an example, to find its partial derivative to a certain $\theta$,
\begin{equation}
\setlength{\abovedisplayskip}{3pt}
\frac{\partial RY(\theta)}{\partial \theta} =(-i\frac{1}{2}Y) RY(\theta)
\setlength{\belowdisplayskip}{3pt}
\end{equation}
$f(\theta)$ becomes
\begin{equation}
\setlength{\abovedisplayskip}{3pt}
    f(\theta)=\left\langle \psi\left\lvert RY^{\dagger }(\theta) \hat{O} RY(\theta)\right\lvert \psi\right\rangle
\end{equation}
Taking the partial derivative of $f(\theta)$, through rigorous mathematical deduction, it can be concluded that
\begin{equation}
\begin{aligned}
    \frac{\partial f(\theta)}{\partial \theta} &= \frac{1}{2} \left[f\left(\theta+\frac{\pi}{2}\right)-f\left(\theta-\frac{\pi}{2}\right)\right]
\end{aligned}\label{eq20}
\end{equation}

The above Eq. \eqref{eq20} seems to be just a differential form, but this shift is a macroscopic variable such as $\frac{\pi }{2}$, so the gradient is exact. It opens up the possibility of using many optimization methods involving gradients in quantum machine learning.

\section{The PID system in BESIII}\label{sec3}
\begin{figure}[h]
    \centering
    \subfigure[]{
		\includegraphics[width=0.46\linewidth]{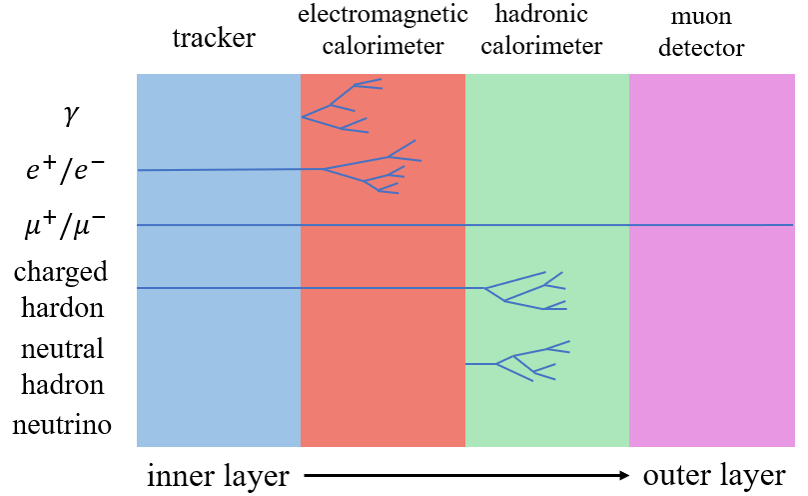}}
	\subfigure[]{
 		\raisebox{5pt}{\includegraphics[width=0.496\linewidth]{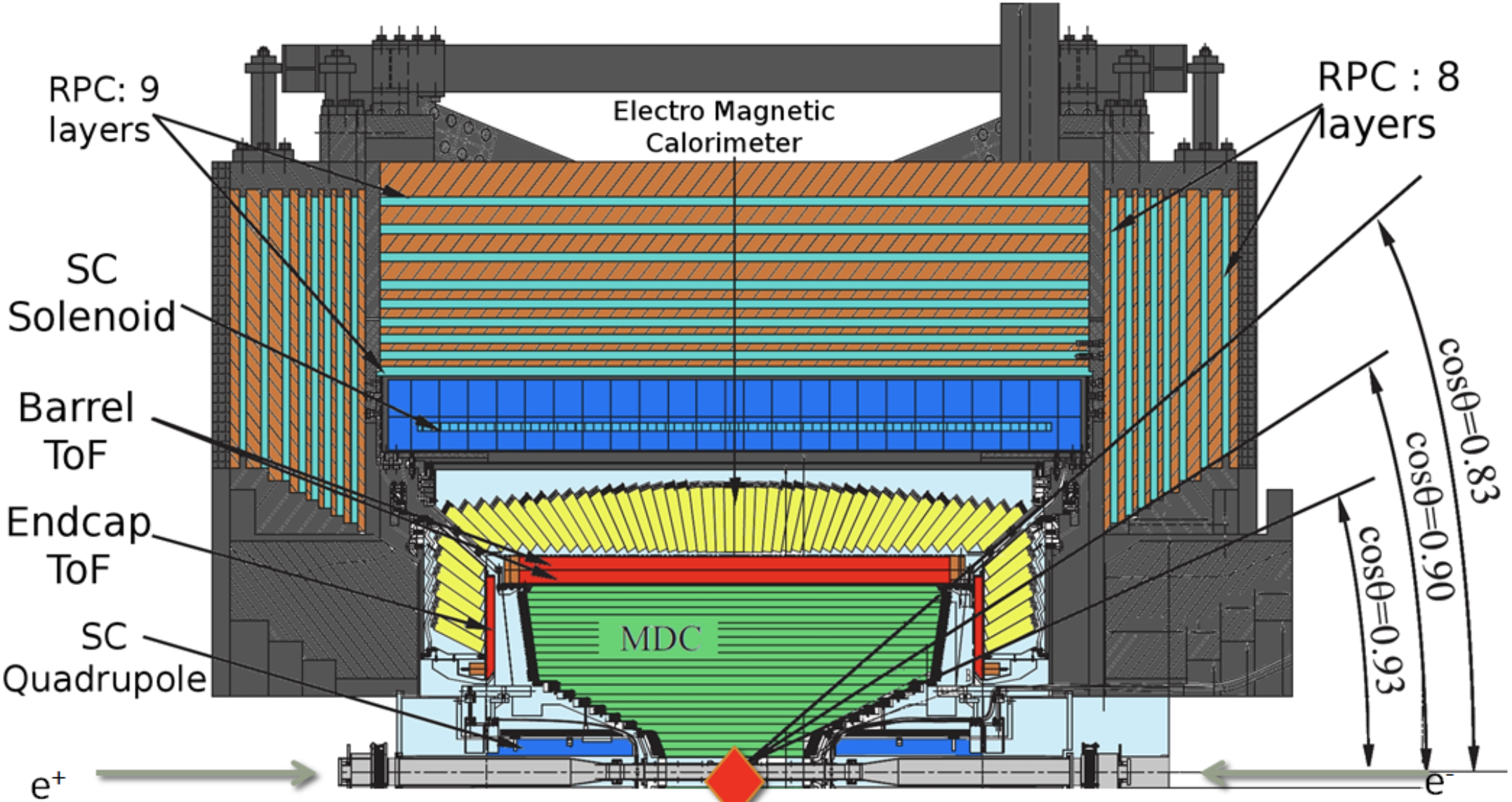}}}
   \caption{(a) The interactions between various particles and different detectors.  (b) The Schematic drawing of the BESIII detector (the upper half)}
   \label{Fig2}
\end{figure}

As indicated in Fig. \ref{Fig2}(a), PID is usually a task to identify the signatures of different species of tracks in various sub-systems.
Due to the absence of a hadronic calorimeter, the discrimination between $\mu^{\pm}$ and $\pi^{\pm}$ is the most difficult PID problem for BESIII.
Therefore, for the $\mu^{\pm}  /\pi^{\pm}$ discrimination, combining information from multiple sub-systems is necessary. 

From the inner layer to the outer layer, Fig. \ref{Fig2}(b) shows the design of the BESIII detector that consists of a main drift chamber (MDC), a time of flight counter (TOF), a CsI(Tl) electromagnetic calorimeter (EMC), a superconducting solenoid magnet and a muon counter (MUC).

The BESIII detector provides the following information for PID:

\begin{itemize}
\item The ionization energy loss of charged particles per unit pathlength in the MDC ($dE/dx$). The MDC consists of 43 layers of sensitive wires, filled with a helium-based gas (60:40 mixture of He and C$_{3}$H$_{8}$) that can be ionized by charged particles passing through. The normalized pulse height of the readout, proportional to the energy loss of the charged particles is a function of $\beta\gamma  =p/m$, where $p$ and $m$ are the momentum and mass of the particle. Combined with the trajectory and momentum measured by the MDC, the $dE/dx$ can better separate charged particles (e, K, and p).

\item The time of flight measured by the TOF counter combined with trajectory and momentum measured by the MDC provides additional discrimination power for particles with different masses.

\item The shape of electromagnetic or hadronic showers in the EMC. The EMC consists of 6240 large CsI(Tl) crystals that can detect electromagnetic showers produced by incident particles. Since the electromagnetic shower shape is distinctive for electrons, muons, and hadrons, the EMC offers some discrimination power for these particles.

\item Hit pattern in the MUC. The muon system of BESIII consists of nine layers of resistive plate chambers (RPC) in the steel magnetic flux return. Most of the hadrons passed the EMC will be absorbed by the return irons. The muons, with strong punching ability, usually leave one hit on each layer of the RPC, while punch-through pions tend to leave multiple hits on the layer where strong interaction occurs. This gives the main discrimination ability between muons and charged hadrons.
\end{itemize}

\section{Muon/Pion Discrimination at BESIII}\label{sec4}
\subsection{Data Sample}
The training and testing data samples are simulated using the BesEvtGen generator \cite{bib15} under the BOSS software \cite{bib16} of BESIII.  To reduce statistical fluctuation, ten independent datasets are selected from a large number of simulated samples, each containing 200 tracks for training and 200 tracks for testing, with the balanced number of $\mu^{\pm}$ and $\pi^{\pm}$. 
To obtain the best $\mu^{\pm}/\pi^{\pm}$ discrimination performance, the reconstructed information of all four sub-detectors is used. The selected features include the following:

\begin{figure}[h]
\centering
\includegraphics[scale=0.55]{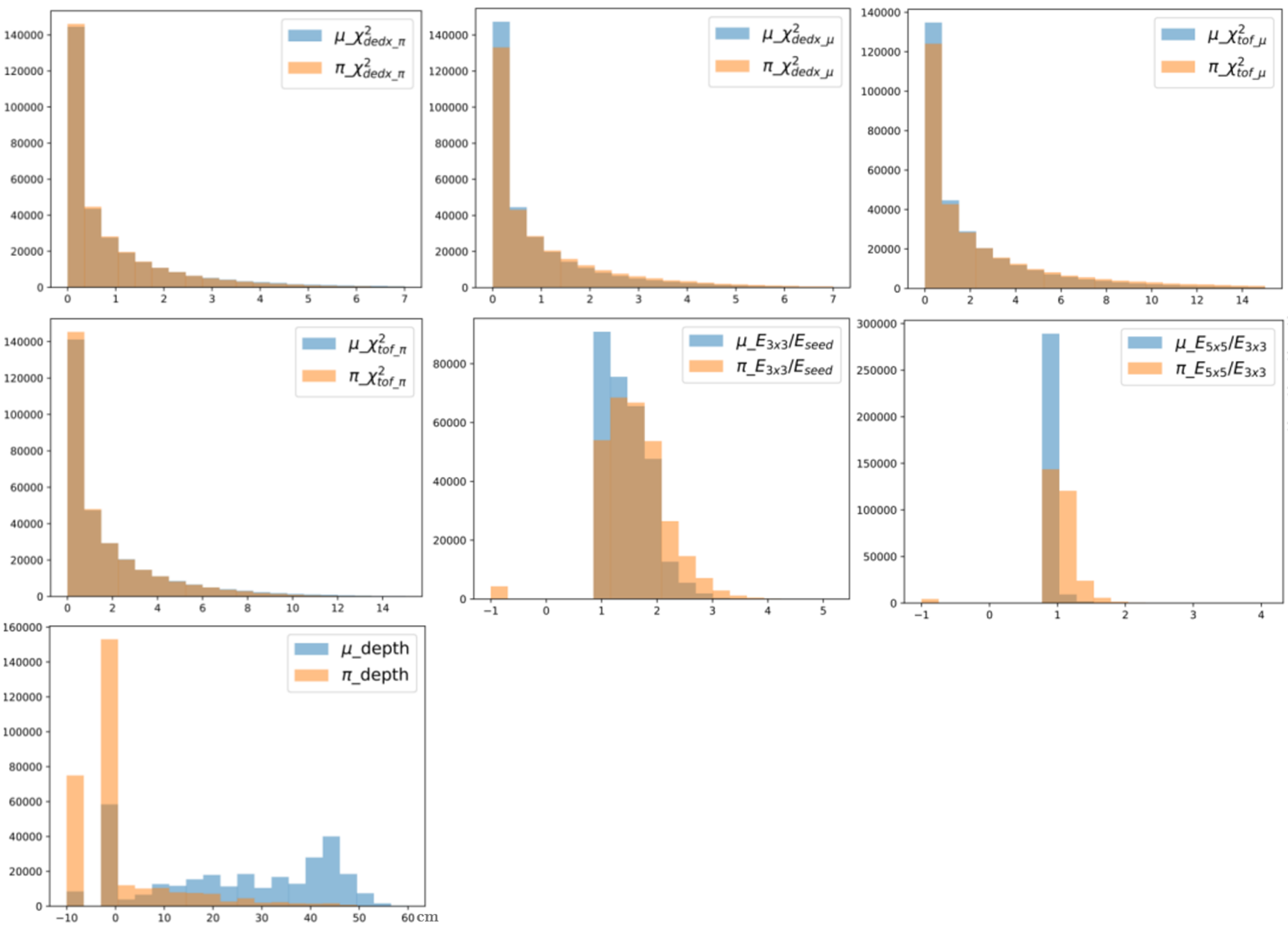}
\caption{Distributions of the selected features before normalization. To provide a clean separation, for tracks that don't reach the EMC or are not successfully reconstructed by the EMC, the $E_{3\times 3}/E_{seed}$ and $E_{5\times 5}/E_{3\times 3}$ are set to be -1. Similar is done for the MUC, where the depth is set to be -10}
\label{Fig3}
\end{figure}

\begin{itemize}
\item The momentum $p$ and the incident angle $\theta$ of the reconstructed track. To minimize the model bias, the $p$ and $\theta$ distributions are flattened. 

\item $\chi^{2}_{\mu\_dedx}$ and $\chi^{2}_{\pi\_dedx}$ given by the MDC. According to \cite{bib17}, the expected $dE/dx$ is a function of $\beta \gamma = p/m$. Therefore, based on the maximum likelihood method the $\chi$ values for $\mu^{\pm}/\pi^{\pm}$ can be calculated as 
\begin{eqnarray}
\hspace{-3mm}
\chi_{i\_dedx} = \frac{dE/dx_{meas} - dE/dx_{exp}}{\sigma_{dedx}} (i=\mu,\pi)
\end{eqnarray}
The $dE/dx_{meas}$ is the measured $dE/dx$ value, while the $dE/dx_{exp}$ is the expected $dE/dx$ under different hypotheses of particle species. The $\sigma_{dedx}$ is the $dE/dx$ resolution.

\item $\chi^{2}_{\mu\_tof}$ and $\chi^{2}_{\pi\_tof}$ given by the TOF. Similar with the $dE/dx$, the $\chi$ values given by TOF is calculated as:
\begin{eqnarray}
\chi_{i\_tof} = \frac{t_{meas} - t_{exp}}{\sigma_{tof}} && (i=\mu,\pi)
\end{eqnarray}

The $t_{meas}$ is the time of flight measured by the TOF counter, while the $t_{exp}$ is the expected time of flight under each particle hypothesis, and the $\sigma_{tof}$ is the time resolution.

\item The ratios of the energy deposit in the EMC, $E_{3\times 3}/E_{seed}$ and $E_{5\times 5}/E_{3\times 3}$. The $E_{seed}$ is the energy deposit in the seed crystal of the shower, while the $E_{3\times 3}$ and $E_{5\times 5}$ is the sum of the energy deposit in the $3\times 3$ and $5\times 5$ crystal array surrounding the seed crystal. 

\item The penetration depth in the MUC in terms of centimeters.

\end{itemize}

Fig. \ref{Fig3} shows the distributions of the features of the data sample. 

\begin{figure}[h]
\centering
\includegraphics[scale=0.65]{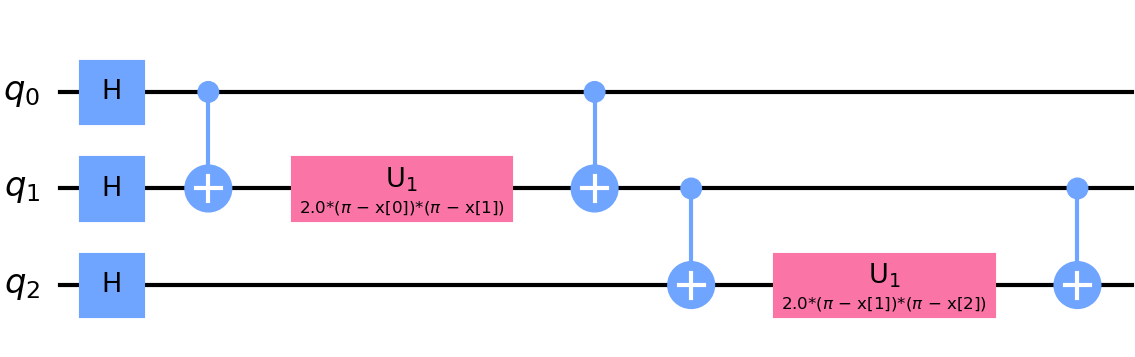}
\caption{Feature map with the Pauli matrix [‘ZZ’]. The U1-gate performs a rotation around the Z-axis direction}
\label{Fig4}
\end{figure}

\begin{table}[h]
\caption{AUCs of various feature maps}
\label{table1}
\begin{tabular*}{\textwidth}{@{\extracolsep\fill}ccccc}
\toprule%
\multicolumn{1}{l}{Circuit\footnotemark[1]} & \multicolumn{1}{l}{Repetitions\footnotemark[2]} & \multicolumn{1}{l}{Entanglement\footnotemark[3]} & EfficientSU2        & RealAmplitudes      \\ \midrule
\multirow{2}{*}{X}           & 2                                & \multirow{2}{*}{none}                                  & 0.839$\pm $0.032 & 0.830$\pm $0.033 \\
                             & 3                                &                                                        & \resizebox{0.13\textwidth}{!}{\textbf{0.851$\pm $0.027}} &  0.846$\pm $0.030 \\ \hline
\multirow{4}{*}{XX}          & \multirow{2}{*}{2}               & linear                                                 & 0.621$\pm $0.038 & 0.624$\pm $0.048 \\
                             &                                  & full                                                   & 0.657$\pm $0.027 & 0.615$\pm $0.035 \\
                             & \multirow{2}{*}{3}               & linear                                                 & 0.621$\pm $0.058 & 0.609$\pm $0.060 \\
                             &                                  & full                                                   & 0.533$\pm $0.056 & 0.528$\pm $0.627 \\ \midrule
\multirow{3}{*}{Z}           & 1                                & \multirow{3}{*}{none}                                  & 0.831$\pm $0.030 & 0.814$\pm $0.058 \\
                             & 2                                &                                                        &  0.845$\pm $0.036 & \resizebox{0.13\textwidth}{!}{\textbf{0.847$\pm $0.027}} \\
                             & 3                                &                                                        & 0.757$\pm $0.023 & 0.691$\pm $0.076 \\ \midrule
\multirow{6}{*}{ZZ}          & \multirow{2}{*}{1}               & linear                                                 & 0.615$\pm $0.068 & 0.613$\pm $0.051 \\
                             &                                  & full                                                   & 0.617$\pm $0.049 & 0.618$\pm $0.061 \\
                             & \multirow{2}{*}{2}               & linear                                                 & 0.620$\pm $0.022 & 0.620$\pm $0.041 \\
                             &                                  & full                                                   & 0.535$\pm $0.048 & 0.546$\pm $0.033 \\
                             & \multirow{2}{*}{3}               & linear                                                 & 0.590$\pm $0.035 & 0.598$\pm $0.036 \\
                             &                                  & full                                                   & 0.518$\pm $0.532 & 0.501$\pm $0.037 \\ 
\botrule                        
\end{tabular*}
\footnotetext[1]{The Pauli matrices used in the Pauli Expansion circuit.}
\footnotetext[2]{The number of repetitions for a feature map.}
\footnotetext[3]{The entanglement method between qubits, where full represents entanglement between any two qubits, and linear represents entanglement only between adjacent qubits.}
\end{table}

\subsection{Result}
In this study, the IBM quantum simulator based on \texttt{qiskit} 0.30.0 and  \texttt{qiskit\_machine\_learning} 0.2.1 are used to perform the following simulation experiments. Based on two typical variational ansatzes in \texttt{qiskit}, EfficientSU2 and RealAmplitudes, a broad series of feature maps built with rotation gates and Controlled-Not (CNOT) gates are simulated to find an optimal circuit structure to encode the data. The EfficientSU2 ansatz consists of single-qubit operation layers spanned by SU(2) and CNOT entanglements, where SU(2) stands for 2×2 unitary matrices, e.g. the Pauli rotation gates $RY$ and $RZ$. The RealAmplitudes has a similar structure to EfficientSU2, with the same CNOT entanglement gates, but it uses only $RY$ to construct the rotation layer, which eliminates the complex part of the prepared quantum state. The performance is the average AUC value obtained by testing these circuits on 10 datasets, as shown in Table \ref{table1}. Fig. \ref{Fig4} visualizes a typical data encoding circuit with rotation gates and entanglement gates on three qubits. 

\begin{figure}[h]
    \centering
    \subfigure[]{
		\includegraphics[width=0.46\linewidth]{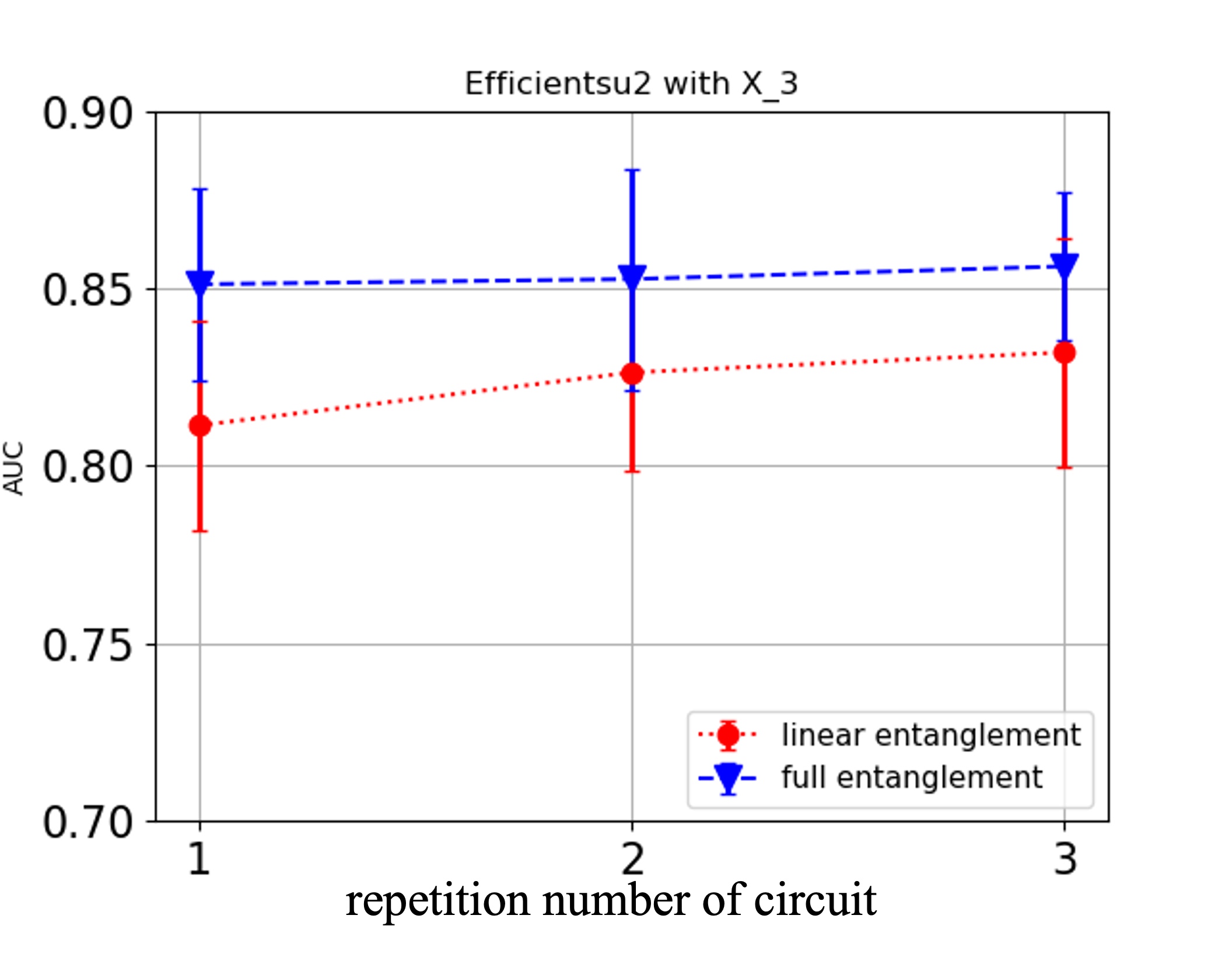}}
	\subfigure[]{
 	\includegraphics[width=0.46\linewidth]{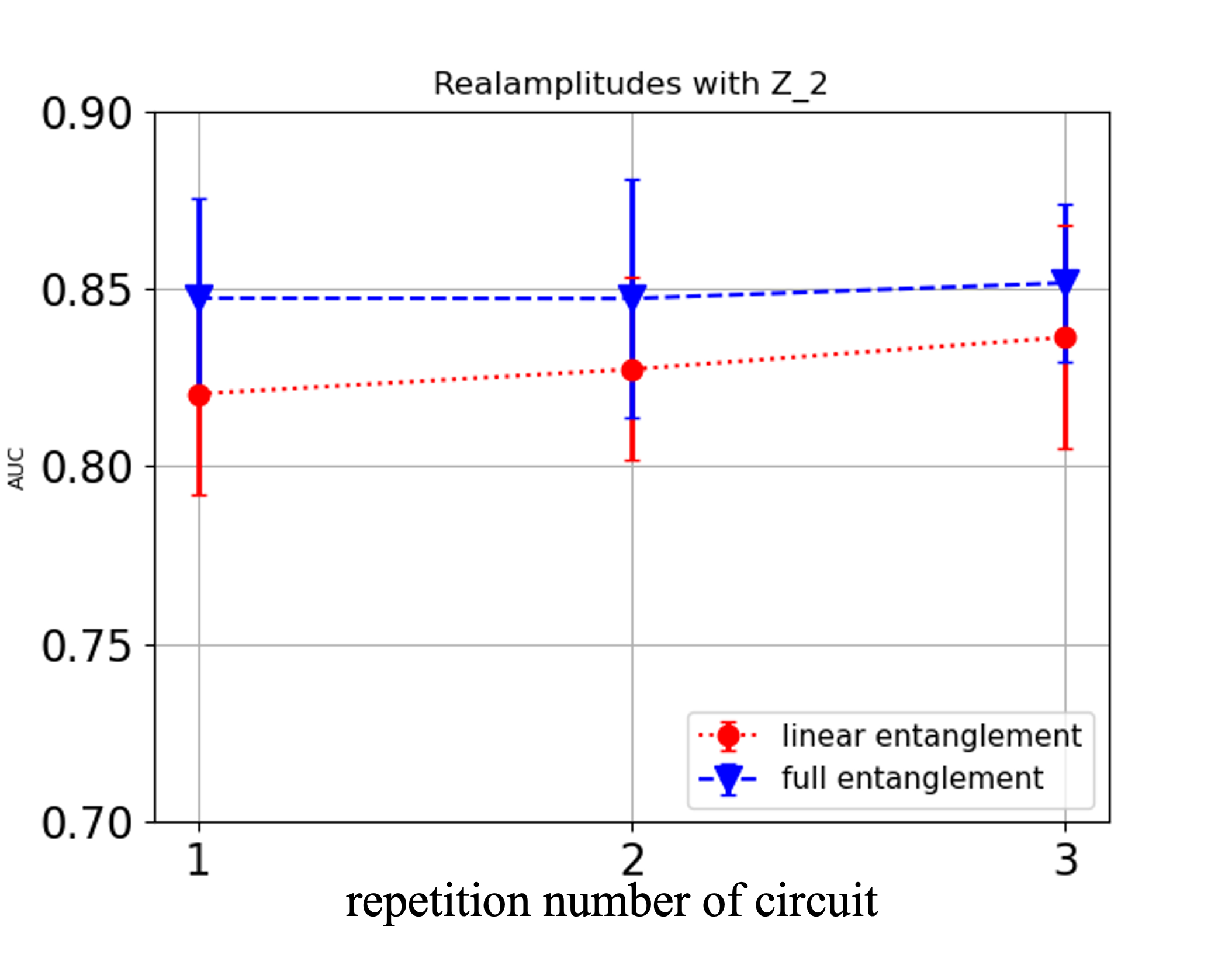}}
   \caption{\centering Effect of variational circuit structure on AUC}
   \label{Fig5}
\end{figure}

As shown in Table \ref{table1}, the performance of different feature maps varies significantly. Relatively simpler circuits built with Pauli rotation gates provide much stronger discrimination power, while circuits with complex entanglement structures show poor discrimination performance. In the subsequent experiments, the corresponding two optimal feature maps were selected for EfficientSU2 and RealAmplitudes, namely circuit X with 3 repetitions ($X\_3$) and circuit Z with 2 repetitions ($Z\_2$).

\begin{figure}[h]
\centering
\subfigure[]{\includegraphics[width=0.46\textwidth]{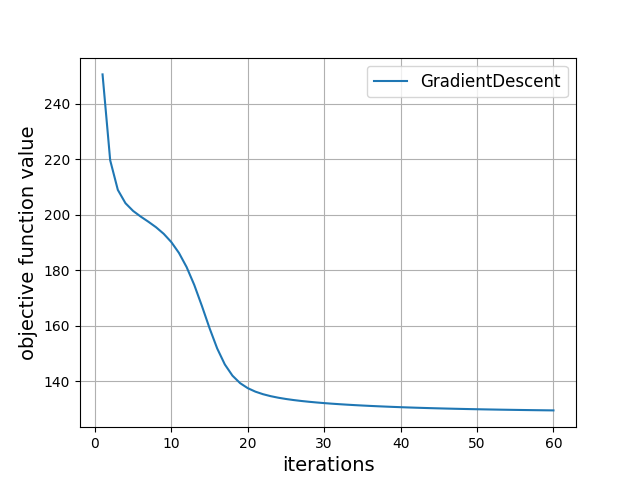}}
\subfigure[]{\includegraphics[width=0.46\textwidth]{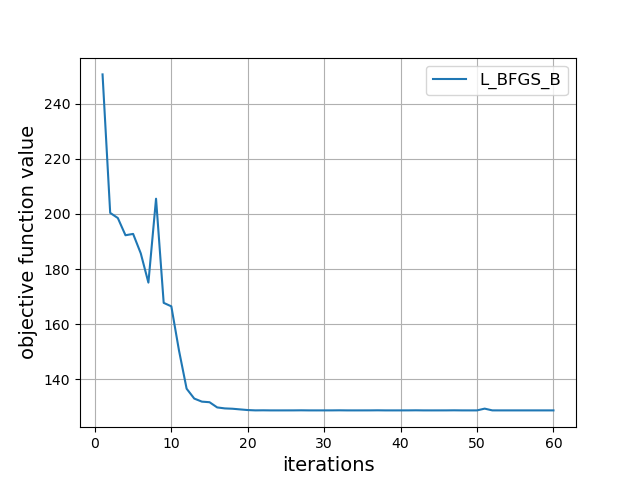}}\\
\subfigure[]{\includegraphics[width=0.46\textwidth]{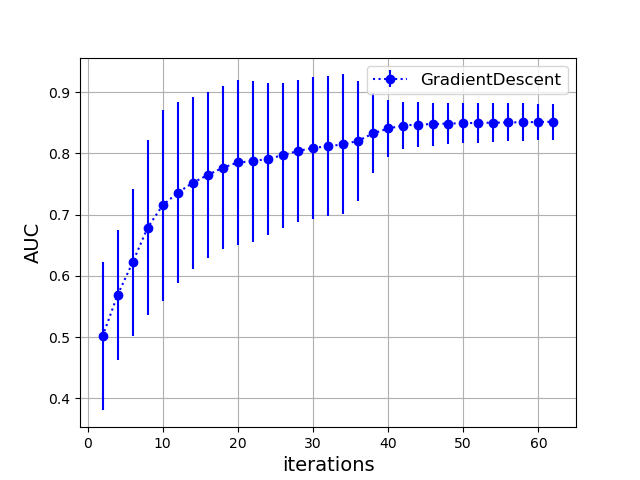}}
\subfigure[]{\includegraphics[width=0.46\textwidth]{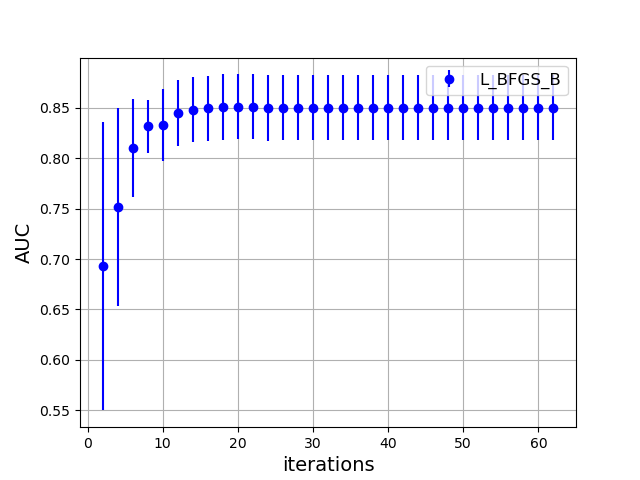}}
\caption{(a)(b) Variation in the objective function during the iteration (c)(d) Variation of AUC with the number of iterations}
\label{Fig6}
\end{figure} 

To further evaluate the impact of the complexity of the ansatzes, two types of ansatzes that best fit $X\_3$ and $Z\_2$ are studied. As shown in Fig. \ref{Fig5}, regardless of the structure of the ansatzes, the full entanglement method consistently outperforms the linear entanglement method. Moreover, the increasing depth of the variational circuits brings more trainable parameters, which allow the model to better capture the latent feature within the dataset, thus the discrimination power of the model is further improved. Although increasing the depth of the variational circuits can improve the performance of the model, on real quantum computing hardware, depth can be limited by noise and decoherence, which may hurt the performance. Therefore, it is important to strike a balance between hardware limitations and model performance in practical applications.

As introduced in Sect. \ref{sec2.4}, the optimization of the trainable parameters associated with the variational circuits is one of the most critical issues in VQC. The classical gradient descent as well as the L\_BFGS\_B(quasi-Newton method) \cite{bib18} algorithm introduced in Sect. \ref{sec2.4} are tested. Figure \ref{Fig6}(a)(b) shows learning curves during the optimization, indicating VQC does have the ability to continuously extract critical information from the data. These two optimizers show quite comparable abilities in finding the optimal solution. However, as shown in Fig. \ref{Fig6}(c)(d), the  L\_BFGS\_B algorithm can converge much faster than the gradient descent algorithm, as expected. 

\begin{figure}[h]
\centering
\includegraphics[width=0.8\textwidth]{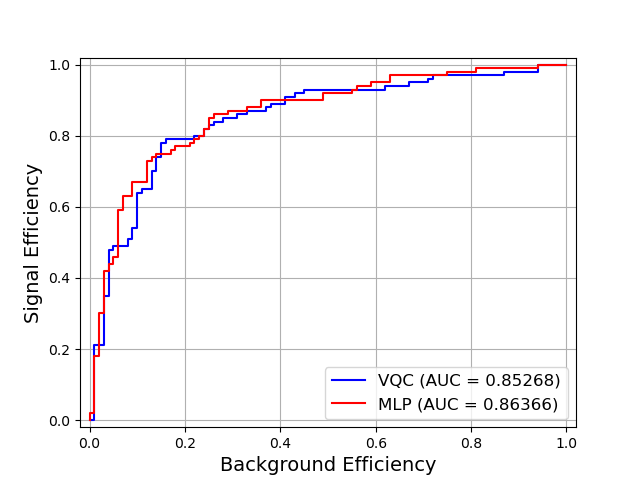}
\caption{\centering The ROC curves of VQC(EfficientSU2 with X\_3 and L\_BFGS\_B) and MLP }
\label{Fig7}
\end{figure}

To compare the performance of VQC and its classical counterpart, a multilayer perceptron (MLP) neural network is trained with the same dataset and features, and their discrimination power is compared. The hyper-parameters of MLP with 4 fully connected layers (number of neurons: 128-64-32-2) are tuned based on grid search, while the VQC is built by combining the optimal configurations from the above experiments. Fig. \ref{Fig7} shows the comparison between the MLP neural network and the VQC, showing that the VQC performs similarly to the classical neural network. 

\section{Summary}
Based on the BESIII MC sample, we developed a VQC model on the quantum simulator for the $\mu^{\pm}  /\pi^{\pm}$  discrimination problem at the BESIII detector. In our study, we first searched expansively for the feature maps to evaluate their performance of encoding classical data into quantum states. The results show that feature maps with simpler structures tend to perform better than those with complex entanglements. We also investigated the specific structure of the variational ansatzes, such as the entanglement method and repetitions, on the performance of the model. For the $\mu^{\pm}  /\pi^{\pm}$ discrimination problem at BESIII, structures that contain complex entanglement or deeper variational circuits tend to improve the performance of the model, but at the cost of more computational time.  However, the performance of different circuit structures depends heavily on the specific application, so they need to be carefully tuned for each problem. The experiments also show that classical optimizers such as gradient descent and the quasi-Newton method can have a large impact on the converging speed of the Variational Hybrid Quantum-Classical Algorithm. By comparing the performance of VQC and its classical counterpart, the MLP neural network, we found they exhibit similar performance on small datasets.

As an exploration of applying quantum machine learning to HEP experiments, this study shows the possibility of applying quantum computing to the processing of real physical data. With the development of quantum hardware in the future, more applications can be foreseen.

\section{Acknowledgements}
This work was supported by National Natural Science Foundation of China (NSFC) under Contracts Nos. 12025502, 12105158, 12188102.

\section{Data Availability Statement}
All data that support the plots and findings within this paper are available from the corresponding author upon reasonable request.

\bibliography{sn-bibliography}% common bib file


\begin{thebibliography}{23}
\ifx \bisbn   \undefined \def \bisbn  #1{ISBN #1}\fi
\ifx \binits  \undefined \def \binits#1{#1}\fi
\ifx \bauthor  \undefined \def \bauthor#1{#1}\fi
\ifx \batitle  \undefined \def \batitle#1{#1}\fi
\ifx \bjtitle  \undefined \def \bjtitle#1{#1}\fi
\ifx \bvolume  \undefined \def \bvolume#1{\textbf{#1}}\fi
\ifx \byear  \undefined \def \byear#1{#1}\fi
\ifx \bissue  \undefined \def \bissue#1{#1}\fi
\ifx \bfpage  \undefined \def \bfpage#1{#1}\fi
\ifx \blpage  \undefined \def \blpage #1{#1}\fi
\ifx \burl  \undefined \def \burl#1{\textsf{#1}}\fi
\ifx \doiurl  \undefined \def \doiurl#1{\url{https://doi.org/#1}}\fi
\ifx \betal  \undefined \def \betal{\textit{et al.}}\fi
\ifx \binstitute  \undefined \def \binstitute#1{#1}\fi
\ifx \binstitutionaled  \undefined \def \binstitutionaled#1{#1}\fi
\ifx \bctitle  \undefined \def \bctitle#1{#1}\fi
\ifx \beditor  \undefined \def \beditor#1{#1}\fi
\ifx \bpublisher  \undefined \def \bpublisher#1{#1}\fi
\ifx \bbtitle  \undefined \def \bbtitle#1{#1}\fi
\ifx \bedition  \undefined \def \bedition#1{#1}\fi
\ifx \bseriesno  \undefined \def \bseriesno#1{#1}\fi
\ifx \blocation  \undefined \def \blocation#1{#1}\fi
\ifx \bsertitle  \undefined \def \bsertitle#1{#1}\fi
\ifx \bsnm \undefined \def \bsnm#1{#1}\fi
\ifx \bsuffix \undefined \def \bsuffix#1{#1}\fi
\ifx \bparticle \undefined \def \bparticle#1{#1}\fi
\ifx \barticle \undefined \def \barticle#1{#1}\fi
\bibcommenthead
\ifx \bconfdate \undefined \def \bconfdate #1{#1}\fi
\ifx \botherref \undefined \def \botherref #1{#1}\fi
\ifx \url \undefined \def \url#1{\textsf{#1}}\fi
\ifx \bchapter \undefined \def \bchapter#1{#1}\fi
\ifx \bbook \undefined \def \bbook#1{#1}\fi
\ifx \bcomment \undefined \def \bcomment#1{#1}\fi
\ifx \oauthor \undefined \def \oauthor#1{#1}\fi
\ifx \citeauthoryear \undefined \def \citeauthoryear#1{#1}\fi
\ifx \endbibitem  \undefined \def \endbibitem {}\fi
\ifx \bconflocation  \undefined \def \bconflocation#1{#1}\fi
\ifx \arxivurl  \undefined \def \arxivurl#1{\textsf{#1}}\fi
\csname PreBibitemsHook\endcsname

%%% 1
\bibitem[\protect\citeauthoryear{Ablikim et~al.}{2010}]{bib1}
\begin{barticle}
\bauthor{\bsnm{Ablikim}, \binits{M.}},
\bauthor{\bsnm{An}, \binits{Z.}},
\bauthor{\bsnm{Bai}, \binits{J.}},
\bauthor{\bsnm{Berger}, \binits{N.}},
\bauthor{\bsnm{Bian}, \binits{J.}},
\bauthor{\bsnm{Cai}, \binits{X.}},
\bauthor{\bsnm{Cao}, \binits{G.}},
\bauthor{\bsnm{Cao}, \binits{X.}},
\bauthor{\bsnm{Chang}, \binits{J.}},
\bauthor{\bsnm{Chen}, \binits{C.}}, \betal:
\batitle{Design and construction of the besiii detector}.
\bjtitle{Nuclear Instruments and Methods in Physics Research Section A: Accelerators, Spectrometers, Detectors and Associated Equipment}
\bvolume{614}(\bissue{3}),
\bfpage{345}--\blpage{399}
(\byear{2010})
\doiurl{10.1016/j.nima.2009.12.050}
\end{barticle}
\endbibitem

%%% 2
\bibitem[\protect\citeauthoryear{Radovic et~al.}{2018}]{radovic2018machine}
\begin{barticle}
\bauthor{\bsnm{Radovic}, \binits{A.}},
\bauthor{\bsnm{Williams}, \binits{M.}},
\bauthor{\bsnm{Rousseau}, \binits{D.}},
\bauthor{\bsnm{Kagan}, \binits{M.}},
\bauthor{\bsnm{Bonacorsi}, \binits{D.}},
\bauthor{\bsnm{Himmel}, \binits{A.}},
\bauthor{\bsnm{Aurisano}, \binits{A.}},
\bauthor{\bsnm{Terao}, \binits{K.}},
\bauthor{\bsnm{Wongjirad}, \binits{T.}}:
\batitle{Machine learning at the energy and intensity frontiers of particle physics}.
\bjtitle{Nature}
\bvolume{560}(\bissue{7716}),
\bfpage{41}--\blpage{48}
(\byear{2018})
\doiurl{10.1038/s41586-018-0361-2}
\end{barticle}
\endbibitem

%%% 3
\bibitem[\protect\citeauthoryear{Feickert and Nachman}{2021}]{feickert2021living}
\begin{barticle}
\bauthor{\bsnm{Feickert}, \binits{M.}},
\bauthor{\bsnm{Nachman}, \binits{B.}}:
\batitle{A living review of machine learning for particle physics}.
\bjtitle{arXiv preprint arXiv:2102.02770}
(\byear{2021})
\doiurl{10.48550/arXiv.2102.02770}
\end{barticle}
\endbibitem

%%% 4
\bibitem[\protect\citeauthoryear{Roe et~al.}{2005}]{bib2}
\begin{barticle}
\bauthor{\bsnm{Roe}, \binits{B.P.}},
\bauthor{\bsnm{Yang}, \binits{H.-J.}},
\bauthor{\bsnm{Zhu}, \binits{J.}},
\bauthor{\bsnm{Liu}, \binits{Y.}},
\bauthor{\bsnm{Stancu}, \binits{I.}},
\bauthor{\bsnm{McGregor}, \binits{G.}}:
\batitle{Boosted decision trees as an alternative to artificial neural networks for particle identification}.
\bjtitle{Nuclear Instruments and Methods in Physics Research Section A: Accelerators, Spectrometers, Detectors and Associated Equipment}
\bvolume{543}(\bissue{2-3}),
\bfpage{577}--\blpage{584}
(\byear{2005})
\doiurl{10.1016/j.nima.2004.12.018}
\end{barticle}
\endbibitem

%%% 5
\bibitem[\protect\citeauthoryear{Bishop et~al.}{1995}]{bib3}
\begin{bbook}
\bauthor{\bsnm{Bishop}, \binits{C.M.}}, \betal:
\bbtitle{Neural Networks for Pattern Recognition}.
\bpublisher{Oxford university press},
\blocation{New York}
(\byear{1995})
\end{bbook}
\endbibitem

%%% 6
\bibitem[\protect\citeauthoryear{Biamonte et~al.}{2017}]{bib4}
\begin{barticle}
\bauthor{\bsnm{Biamonte}, \binits{J.}},
\bauthor{\bsnm{Wittek}, \binits{P.}},
\bauthor{\bsnm{Pancotti}, \binits{N.}},
\bauthor{\bsnm{Rebentrost}, \binits{P.}},
\bauthor{\bsnm{Wiebe}, \binits{N.}},
\bauthor{\bsnm{Lloyd}, \binits{S.}}:
\batitle{Quantum machine learning}.
\bjtitle{Nature}
\bvolume{549}(\bissue{7671}),
\bfpage{195}--\blpage{202}
(\byear{2017})
\doiurl{10.1038/nature23474}
\end{barticle}
\endbibitem

%%% 7
\bibitem[\protect\citeauthoryear{Schuld and Petruccione}{2018}]{bib5}
\begin{bbook}
\bauthor{\bsnm{Schuld}, \binits{M.}},
\bauthor{\bsnm{Petruccione}, \binits{F.}}:
\bbtitle{Supervised Learning with Quantum Computers}
vol. \bseriesno{17}.
\bpublisher{Springer},
\blocation{Berlin}
(\byear{2018})
\end{bbook}
\endbibitem

%%% 8
\bibitem[\protect\citeauthoryear{Guan et~al.}{2021}]{guan2021quantum}
\begin{barticle}
\bauthor{\bsnm{Guan}, \binits{W.}},
\bauthor{\bsnm{Perdue}, \binits{G.}},
\bauthor{\bsnm{Pesah}, \binits{A.}},
\bauthor{\bsnm{Schuld}, \binits{M.}},
\bauthor{\bsnm{Terashi}, \binits{K.}},
\bauthor{\bsnm{Vallecorsa}, \binits{S.}},
\bauthor{\bsnm{Vlimant}, \binits{J.-R.}}:
\batitle{Quantum machine learning in high energy physics}.
\bjtitle{Machine Learning: Science and Technology}
\bvolume{2}(\bissue{1}),
\bfpage{011003}
(\byear{2021})
\doiurl{10.1088/2632-2153/abc17d}
\end{barticle}
\endbibitem

%%% 9
\bibitem[\protect\citeauthoryear{Wu and Yoo}{2022}]{wu2022challenges}
\begin{barticle}
\bauthor{\bsnm{Wu}, \binits{S.L.}},
\bauthor{\bsnm{Yoo}, \binits{S.}}:
\batitle{Challenges and opportunities in quantum machine learning for high-energy physics}.
\bjtitle{Nature Reviews Physics}
\bvolume{4}(\bissue{3}),
\bfpage{143}--\blpage{144}
(\byear{2022})
\end{barticle}
\endbibitem

%%% 10
\bibitem[\protect\citeauthoryear{Rebentrost et~al.}{2014}]{bib6}
\begin{barticle}
\bauthor{\bsnm{Rebentrost}, \binits{P.}},
\bauthor{\bsnm{Mohseni}, \binits{M.}},
\bauthor{\bsnm{Lloyd}, \binits{S.}}:
\batitle{Quantum support vector machine for big data classification}.
\bjtitle{Physical review letters}
\bvolume{113}(\bissue{13}),
\bfpage{130503}
(\byear{2014})
\doiurl{10.1103/PhysRevLett.113.130503}
\end{barticle}
\endbibitem

%%% 11
\bibitem[\protect\citeauthoryear{Schuld et~al.}{2020}]{bib7}
\begin{barticle}
\bauthor{\bsnm{Schuld}, \binits{M.}},
\bauthor{\bsnm{Bocharov}, \binits{A.}},
\bauthor{\bsnm{Svore}, \binits{K.M.}},
\bauthor{\bsnm{Wiebe}, \binits{N.}}:
\batitle{Circuit-centric quantum classifiers}.
\bjtitle{Physical Review A}
\bvolume{101}(\bissue{3}),
\bfpage{032308}
(\byear{2020})
\doiurl{10.1103/PhysRevA.101.032308}
\end{barticle}
\endbibitem

%%% 12
\bibitem[\protect\citeauthoryear{Wu et~al.}{2021}]{bib8}
\begin{barticle}
\bauthor{\bsnm{Wu}, \binits{S.L.}},
\bauthor{\bsnm{Sun}, \binits{S.}},
\bauthor{\bsnm{Guan}, \binits{W.}},
\bauthor{\bsnm{Zhou}, \binits{C.}},
\bauthor{\bsnm{Chan}, \binits{J.}},
\bauthor{\bsnm{Cheng}, \binits{C.L.}},
\bauthor{\bsnm{Pham}, \binits{T.}},
\bauthor{\bsnm{Qian}, \binits{Y.}},
\bauthor{\bsnm{Wang}, \binits{A.Z.}},
\bauthor{\bsnm{Zhang}, \binits{R.}}, \betal:
\batitle{Application of quantum machine learning using the quantum kernel algorithm on high energy physics analysis at the lhc}.
\bjtitle{Physical Review Research}
\bvolume{3}(\bissue{3}),
\bfpage{033221}
(\byear{2021})
\doiurl{10.1103/PhysRevResearch.3.033221}
\end{barticle}
\endbibitem

%%% 13
\bibitem[\protect\citeauthoryear{Belis et~al.}{2021}]{bib9}
\begin{bchapter}
\bauthor{\bsnm{Belis}, \binits{V.}},
\bauthor{\bsnm{Gonz{\'a}lez-Castillo}, \binits{S.}},
\bauthor{\bsnm{Reissel}, \binits{C.}},
\bauthor{\bsnm{Vallecorsa}, \binits{S.}},
\bauthor{\bsnm{Combarro}, \binits{E.F.}},
\bauthor{\bsnm{Dissertori}, \binits{G.}},
\bauthor{\bsnm{Reiter}, \binits{F.}}:
\bctitle{Higgs analysis with quantum classifiers}.
In: \bbtitle{EPJ Web of Conferences},
vol. \bseriesno{251},
p. \bfpage{03070}
(\byear{2021}).
\doiurl{10.1051/epjconf/202125103070}
\end{bchapter}
\endbibitem

%%% 14
\bibitem[\protect\citeauthoryear{Havl{\'\i}{\v{c}}ek et~al.}{2019}]{bib10}
\begin{barticle}
\bauthor{\bsnm{Havl{\'\i}{\v{c}}ek}, \binits{V.}},
\bauthor{\bsnm{C{\'o}rcoles}, \binits{A.D.}},
\bauthor{\bsnm{Temme}, \binits{K.}},
\bauthor{\bsnm{Harrow}, \binits{A.W.}},
\bauthor{\bsnm{Kandala}, \binits{A.}},
\bauthor{\bsnm{Chow}, \binits{J.M.}},
\bauthor{\bsnm{Gambetta}, \binits{J.M.}}:
\batitle{Supervised learning with quantum-enhanced feature spaces}.
\bjtitle{Nature}
\bvolume{567}(\bissue{7747}),
\bfpage{209}--\blpage{212}
(\byear{2019})
\doiurl{10.1038/s41586-019-0980-2}
\end{barticle}
\endbibitem

%%% 15
\bibitem[\protect\citeauthoryear{De~Boer et~al.}{2005}]{bib11}
\begin{barticle}
\bauthor{\bsnm{De~Boer}, \binits{P.-T.}},
\bauthor{\bsnm{Kroese}, \binits{D.P.}},
\bauthor{\bsnm{Mannor}, \binits{S.}},
\bauthor{\bsnm{Rubinstein}, \binits{R.Y.}}:
\batitle{A tutorial on the cross-entropy method}.
\bjtitle{Annals of operations research}
\bvolume{134}(\bissue{1}),
\bfpage{19}--\blpage{67}
(\byear{2005})
\doiurl{10.1007/s10479-005-5724-z}
\end{barticle}
\endbibitem

%%% 16
\bibitem[\protect\citeauthoryear{Davidon}{1991}]{davidon1991variable}
\begin{barticle}
\bauthor{\bsnm{Davidon}, \binits{W.C.}}:
\batitle{Variable metric method for minimization}.
\bjtitle{SIAM Journal on Optimization}
\bvolume{1}(\bissue{1}),
\bfpage{1}--\blpage{17}
(\byear{1991})
\doiurl{10.1137/0801001}
\end{barticle}
\endbibitem

%%% 17
\bibitem[\protect\citeauthoryear{Baydin et~al.}{2018}]{bib12}
\begin{botherref}
\oauthor{\bsnm{Baydin}, \binits{A.G.}},
\oauthor{\bsnm{Pearlmutter}, \binits{B.A.}},
\oauthor{\bsnm{Radul}, \binits{A.A.}},
\oauthor{\bsnm{Siskind}, \binits{J.M.}}:
Automatic differentiation in machine learning: a survey.
Journal of machine learning research
\textbf{18}
(2018)
\end{botherref}
\endbibitem

%%% 18
\bibitem[\protect\citeauthoryear{Bergholm et~al.}{2018}]{bib13}
\begin{barticle}
\bauthor{\bsnm{Bergholm}, \binits{V.}},
\bauthor{\bsnm{Izaac}, \binits{J.}},
\bauthor{\bsnm{Schuld}, \binits{M.}},
\bauthor{\bsnm{Gogolin}, \binits{C.}},
\bauthor{\bsnm{Alam}, \binits{M.S.}},
\bauthor{\bsnm{Ahmed}, \binits{S.}},
\bauthor{\bsnm{Arrazola}, \binits{J.M.}},
\bauthor{\bsnm{Blank}, \binits{C.}},
\bauthor{\bsnm{Delgado}, \binits{A.}},
\bauthor{\bsnm{Jahangiri}, \binits{S.}}, \betal:
\batitle{Pennylane: Automatic differentiation of hybrid quantum-classical computations}.
\bjtitle{arXiv preprint arXiv:1811.04968}
(\byear{2018})
\doiurl{10.48550/arXiv.1811.04968}
\end{barticle}
\endbibitem

%%% 19
\bibitem[\protect\citeauthoryear{Schuld et~al.}{2019}]{bib14}
\begin{barticle}
\bauthor{\bsnm{Schuld}, \binits{M.}},
\bauthor{\bsnm{Bergholm}, \binits{V.}},
\bauthor{\bsnm{Gogolin}, \binits{C.}},
\bauthor{\bsnm{Izaac}, \binits{J.}},
\bauthor{\bsnm{Killoran}, \binits{N.}}:
\batitle{Evaluating analytic gradients on quantum hardware}.
\bjtitle{Physical Review A}
\bvolume{99}(\bissue{3}),
\bfpage{032331}
(\byear{2019})
\doiurl{10.1103/PhysRevA.99.032331}
\end{barticle}
\endbibitem

%%% 20
\bibitem[\protect\citeauthoryear{Ping}{2008}]{bib15}
\begin{barticle}
\bauthor{\bsnm{Ping}, \binits{R.-G.}}:
\batitle{Event generators at besiii}.
\bjtitle{Chinese Physics C}
\bvolume{32}(\bissue{8}),
\bfpage{599}--\blpage{602}
(\byear{2008})
\doiurl{10.1088/1674-1137/32/8/001}
\end{barticle}
\endbibitem

%%% 21
\bibitem[\protect\citeauthoryear{Li et~al.}{2006}]{bib16}
\begin{bchapter}
\bauthor{\bsnm{Li}, \binits{W.-D.}},
\bauthor{\bsnm{Liu}, \binits{H.-M.}},
\bauthor{\bsnm{Deng}, \binits{Z.}},
\bauthor{\bsnm{He}, \binits{K.}},
\bauthor{\bsnm{He}, \binits{M.}},
\bauthor{\bsnm{Ji}, \binits{X.}},
\bauthor{\bsnm{Jiang}, \binits{L.}},
\bauthor{\bsnm{Li}, \binits{H.}},
\bauthor{\bsnm{Liu}, \binits{C.}},
\bauthor{\bsnm{Ma}, \binits{Q.}}, \betal:
\bctitle{The offline software for the besiii experiment}.
In: \bbtitle{Proceeding of CHEP},
vol. \bseriesno{27}
(\byear{2006})
\end{bchapter}
\endbibitem

%%% 22
\bibitem[\protect\citeauthoryear{Blum et~al.}{2008}]{bib17}
\begin{bbook}
\bauthor{\bsnm{Blum}, \binits{W.}},
\bauthor{\bsnm{Riegler}, \binits{W.}},
\bauthor{\bsnm{Rolandi}, \binits{L.}}:
\bbtitle{Particle Detection with Drift Chambers}.
\bpublisher{Springer},
\blocation{Berlin, Heidelberg}
(\byear{2008})
\end{bbook}
\endbibitem

%%% 23
\bibitem[\protect\citeauthoryear{Zhu et~al.}{1997}]{bib18}
\begin{barticle}
\bauthor{\bsnm{Zhu}, \binits{C.}},
\bauthor{\bsnm{Byrd}, \binits{R.H.}},
\bauthor{\bsnm{Lu}, \binits{P.}},
\bauthor{\bsnm{Nocedal}, \binits{J.}}:
\batitle{Algorithm 778: L-bfgs-b: Fortran subroutines for large-scale bound-constrained optimization}.
\bjtitle{ACM Transactions on mathematical software (TOMS)}
\bvolume{23}(\bissue{4}),
\bfpage{550}--\blpage{560}
(\byear{1997})
\doiurl{10.1145/279232.279236}
\end{barticle}
\endbibitem

\end{thebibliography}

\end{document}